\documentstyle[twoside,epsfig]{article}

\title{Polynomial Observables in the Graph Partitioning Problem}
\author{M.A. Marchisio\\Dipartimento di Fisica, Universit\`a di Trento\\I-38050 Povo, Italy\\ email:marchisi@science.unitn.it }
\date{8 January 2001}
\begin{document}
\parindent=15pt
\baselineskip=13pt
\maketitle
\begin{abstract}
\noindent
Although NP-Complete problems are the most difficult
decisional problems, it's possible to discover  in them polynomial (or easy)
observables. We study the Graph Partitioning Problem showing that it's possible to
recognize in it two correlated polynomial observables. The particular behaviour of one
of them with respect to the connectivity of the graph suggests the presence of a phase
transition in partitionability.
\end{abstract}

\section{Introduction}
\noindent
Decisional problems are essentially divided into two great classes: P or {\it  easy}
problems, which can be solved in polynomial time by deterministic algorithms, and NP
or {\it difficult} problems whose worst instances can't be solved in polynomial time,
unless you have a non-deterministic computer. Of course P $\subseteq$ NP but the
question if P$\neq$ NP is still open. Among the NP problems, the so called NP-Complete
are particularly important because they are the most difficult in NP; in fact all NP
problems can be converted into a NP-Complete problem in polynomial time. If just one
NP-Complete problem could be shown to be in P, that would imply  P=NP. Till now, no
polynomial algorithm for any NP-Complete problem is known; however, recently, a
polynomial observable has been found in the K-SAT problem\cite{zecchina1}, which is
NP-Complete\cite{gar}, suggesting that also other NP-Complete problems can display
computationally easy observables.\\
Another relevant feature of NP-Complete problems (but not only of them) was
found about twenty years ago: they exhibit phase transitions. Important results were
obtained for K-SAT\cite{zecchina2,zecchina3}, Number
Partitioning\cite{mertens1,mertens2} and Graph Partitioning\cite{boettcher}.\\
In this work  the Graph Partitioning Problem (GPP) was applied to random
graphs; two correlated P observables were found in it: the number of redbonds\cite{con1,con2} and the
number of redgraphs; moreover, the number of redgraphs could be put in relation to the
phase transition of the problem.\\
In the next section a review of GPP is sketched, while in section $3$ we describe our
polynomial algorithm and discuss the numerical results.

\section{The Graph Partitioning Problem}
\noindent
A graph $G(V,E)$ is assigned by giving a set $V$ of $N$ points called {\it
vertices}, and  a set $E \subset V \times V$ of $m$ {\it edges} which specify which
pairs of vertices are adjacent in $G(V,E)$: each edge connects two distinct vertices.
The graph is a {\it random graph} if each edge exists with some probability $p$. The
{\it mean connectivity} $\alpha$ is defined as the average number of edges per vertex.
Having $N$ vertices and $m$ edges, the total number of possible edges is
$\frac{N(N-1)}{2}$ so $p = \frac{2m}{N(N-1)}$  and $\alpha$ is simply
$\frac{2m}{N}$.\\
In general, given a graph $G(V,E)$, the Graph Partitioning Problem consists in finding
the partition of set $V$ into two disjoint and  equally sized subsets $V_1$ and $V_2$
such that the number $K$ of edges having one vertex in $V_1$ and the other in $V_2$
(bonds) is minimized; if we find a partition with no bonds at all, we will say that
the graph is {\it partitionable}. In decisional form one can simply ask if, given a
graph $G(V,E)$ and a number $K$, there is a partition such that the number of bonds is
lower than or equal to $K$.\\
We know that a large (giant) cluster appears in random graphs at $\alpha =1$, the so
called {\it percolation threshold}, but the giant cluster's size becomes $N/2$ only at
$\alpha_c=2\ln2 \approx 1.386$\cite{erdos,fu} and here GPP shows a phase
transition\cite{boettcher}: when $\alpha < \alpha_c$ random graphs are partitionable, 
but they become suddenly unpartitionable  for $\alpha > \alpha_c$ and the number of
bonds grows up with $\alpha$ (for fixed values of $N$)\cite{boettcher}. It's
reasonable to suppose that partitions with only one bond, the so called redbond
partition, lie only in a small region around $\alpha_c$.\\
A random graph which has at least one redbond will be called redgraph and the mean
number of redbonds per redgraphs at fixed values of $\alpha$ and $N$ represents the
entropy (that is the number of solutions) of the GPP with $K=1$.\\
Why redbond? This name originated from an electronics problem which is a practical
application of GPP\cite{kirk}: the design of an efficient component made of $N$
circuits equally divided over  two chips and connected by $m$ wires. One  would like
to minimize the time required by informations to propagate through the entire
machinery, and it is known that wires connecting circuits on different chips (the
bonds) slow down remarkably  the propagation: so one has to look for a circuits'
placement which minimizes the number of bonds between the two chips. On the other
hand, another effect might come into play; if there were just one link, the whole
information would go through it overloading the bond: in a little while the entire
component would break down because the bond  burns, after turning red.\\
It's worthwhile to note at the end of this section that, like in electronics, infinite range percolation models can be used to cope with problems in a huge variety of different subjects, for example: the origin of life\cite{kauff1,kauff2}, fluctuations in the stock market\cite{cont} and the breakdown of the internet\cite{bara,cohen}  

\section{The method and the results.}
\begin{figure}[h]
   \centerline{
\epsfig{file=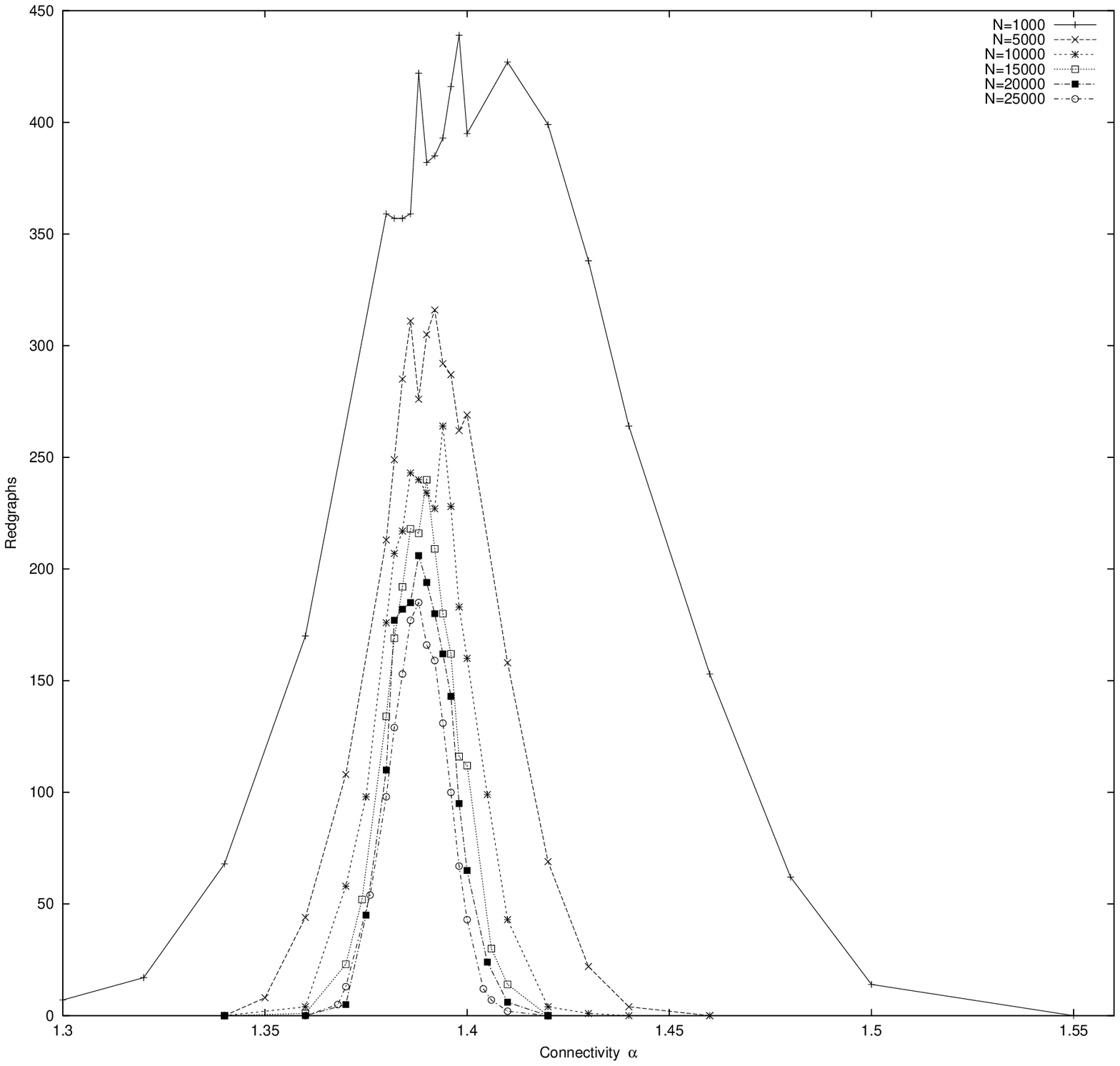}
}
    \caption{Number of redgraphs in terms  of connectivity. The measures are made over
    1000 samples.}
    \label{fig1}
\end{figure}

\noindent 
In order to find a redbond,  after having broken the entire random graph
into clusters (connected subgraphs) using the Hoshen-Kopelman algorithm\cite{hoshen}, the first
step is to find  the giant cluster  and to calculate its size, namely the number of
its vertices. If the giant cluster's size is lower than or equal to $N/2$, the istance
is partitionable and it has no sense to look for a redbond; otherwise, if the size is
greater than $N/2$, the graph could be a redgraph. As a second step,  remove one edge
from the giant cluster: if this edge is a redbond, the stripped giant cluster
separates into two subclusters, the size of the biggest one being lower than
or equal to $N/2$. Putting the just obtained biggest cluster in $V_1$ and the other
subclusters in $V_2$, and completing the partition by inserting into $V_1$ and $V_2$
all other clusters of the original graph, one obtains a partition with a redbond.\\
Therefore, to calculate the number of redbonds, this procedure is applied recursively
to each edge of the giant cluster: one checks  whether the removal of each one leads
to a biggest subcluster with size lower than or equal to $N/2$. Using again the
Hoshen-Kopelman algorithm, the breaking of the giant cluster requires at most $O(N^2)$
steps; since the number of the giant cluster's edges is obviously $O(N^2)$, the
complexity time of the algorithm is $O(N^4)$, so that  the number of redbonds is a
polynomial observable of GPP.\\
If we have a statistical ensemble of $l$ random graphs and  want to know how many
graphs are red, we have only to repeat the above process $l$ times. Therefore,
regarding the number of redgraphs as an observable of the problem, we see that it can
be calculated in $l \; N^4$ steps, so that it is also polynomial.\\ Actually these observables would be better called {\it probable redbonds} and
{\it probable redgraphs}, because, in searching for true redbonds, also the structure
of the non-giant clusters should be taken into account in detail. However, as $N$
increases, the probability to meet a real redbond with our algorithm grows up. We
performed simulations over random graphs with $N=1000$, $5000$, $10000$, $15000$,
$20000$, and $25000$, in order to calculate the number of redgraphs. The range of
$\alpha$ depended on $N$ and varied, for example, from $1.3$ to $1.55$ for $N=1000$
and from $1.36$ to $1.42$ for $N=25000$. For each couple of values of $N$ and $\alpha$
the program crunched $1000$ samples. The outcome is that redgraphs do exist, as
expected, only in a small region around $\alpha_c$, while the probability to find a
redgraph shows a peak close to $\alpha_c$, which becomes smaller and more narrow as
$N$  increases. This behaviour might be interpreted as the signal of the known phase
transition, in analogy to what happens in some physical systems (think  of the
magnetic susceptivity $\chi$ versus temperature in a ferromagnetic system).
(Fig.\ref{fig1}).

\begin{figure}[h]
   \centerline{
\epsfig{file=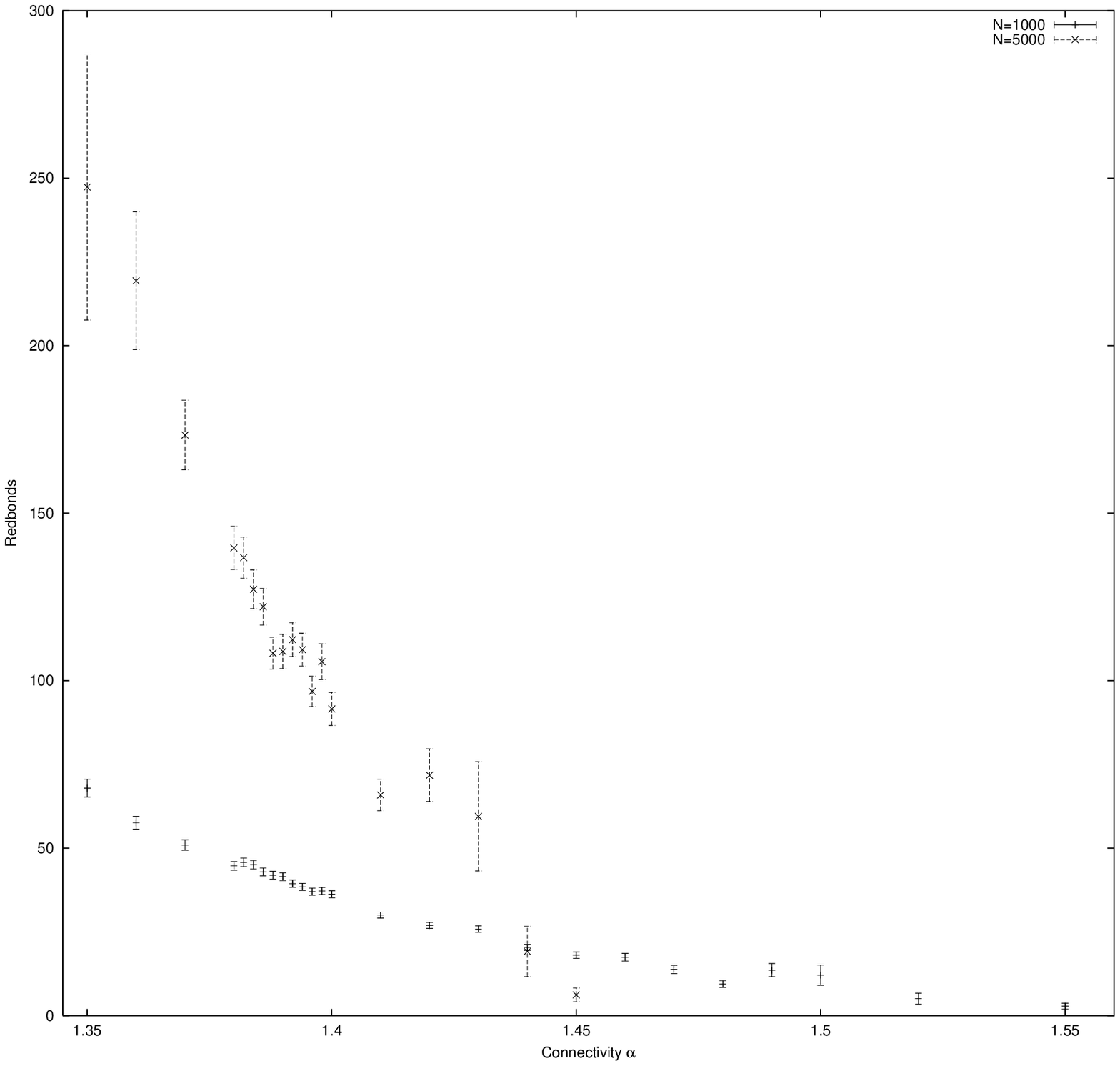}
}
    \caption{Mean number of redbonds (entropy) in terms  of connectivity. The average
    is made over 10000 samples.}
    \label{fig2}
\end{figure}

In order to estimate the entropy, we limited ourselves only to the
cases  $N=1000$ and $N=5000$, taking care to average our results over 10000 samples,
in order to reduce the statistical errors. We verified that entropy doesn't vanish
abruptly at $\alpha_c$, as problem's solutions are found above this point. For
$N=1000$ entropy goes to zero slowly whereas the fall for $N=5000$ is much faster.
This behaviour is a clear  finite-size effect, quite similar to the one  found in the
K-SAT problem\cite {zecchina3}(Fig.\ref{fig2}).

\section{Conclusion}
\noindent
In this paper we have studied the GPP applied to random graphs and we have shown that
two tightly correlated P observables are present in this NP-Complete Problem: the number
of redbonds and the number of redgraphs. We have also characterized in a new way the
phase transition in partitionability of the GPP through the peaked behaviour of the number
of redgraphs near $\alpha_c$.

\section{Acknowledgements}
\noindent
I want to thank G. Ponzano and M. Caselle for their help in doing this work.

\end{document}